\documentclass[sigconf]{acmart}

\AtBeginDocument{%
  \providecommand\BibTeX{{%
    \normalfont B\kern-0.5em{\scshape i\kern-0.25em b}\kern-0.8em\TeX}}}

\setcopyright{acmcopyright}
\copyrightyear{2020}
\acmYear{2020}
\acmDOI{10.1145/1122445.1122456}

\acmConference[XXX '20]{The 26th ACM XXX Conference on Knowledge Discovery and Data Mining}{August 22-27, 2020}{San Diego, CA, USA}
\acmPrice{15.00}
\acmISBN{978-1-4503-XXXX-X/20/08}

\begin{document}

%%
%% The "title" command has an optional parameter,
%% allowing the author to define a "short title" to be used in page headers.
\title{XLBoost-Geo:  An IP Geolocation System Based on Extreme Landmark Boosting}

%%
%% The "author" command and its associated commands are used to define
%% the authors and their affiliations.
%% Of note is the shared affiliation of the first two authors, and the
%% "authornote" and "authornotemark" commands
%% used to denote shared contribution to the research.

\author{Yucheng Wang}
\affiliation{%
  \institution{School of Cyber Security, University of Chinese Academy of Sciences}
  \institution{Beijing Key Laboratory of IOT Information Security Technology, Institute of Information Engineering, Chinese Academy of Sciences}
  \city{Beijing}
  \country{China}
}

\author{Hongsong Zhu}
\authornote{Hongsong Zhu is the corresponding author}
\affiliation{%
  \institution{Beijing Key Laboratory of IOT Information Security Technology, Institute of Information Engineering, Chinese Academy of Sciences}
  \city{Beijing}
  \country{China}
  }
\email{zhuhongsong@iie.ac.cn}

\author{Jinfa Wang}
\affiliation{%
  \institution{Beijing Key Laboratory of IOT Information Security Technology, Institute of Information Engineering, Chinese Academy of Sciences}
  \city{Beijing}
  \country{China}
  }

\author{Jie Liu}
\affiliation{%
  \institution{School of Cyber Security, University of Chinese Academy of Sciences}
  \institution{Beijing Key Laboratory of IOT Information Security Technology, Institute of Information Engineering, Chinese Academy of Sciences}
  \city{Beijing}
  \country{China}
  }

\author{Yong Wang}
\affiliation{%
  \institution{Zhengzhou Aiwen Computer Technology Co. Ltd.}
  \city{Zhengzhou}
  \country{China}
}

\author{Limin Sun}
\affiliation{%
  \institution{School of Cyber Security, University of Chinese Academy of Sciences}
  \institution{Beijing Key Laboratory of IOT Information Security Technology, Institute of Information Engineering, Chinese Academy of Sciences}
  \city{Beijing}
  \country{China}
}

%%
%% By default, the full list of authors will be used in the page
%% headers. Often, this list is too long, and will overlap
%% other information printed in the page headers. This command allows
%% the author to define a more concise list
%% of authors' names for this purpose.
\renewcommand{\shortauthors}{Y. Wang et al.}

%%
%% The abstract is a short summary of the work to be presented in the
%% article.
 \begin{abstract}
% background
IP geolocation aims at locating the geographical position of Internet devices, which plays an essential role in many Internet applications. In this field, a long-standing challenge is how to find a large number of highly-reliable landmarks, which is the key to improve the precision of IP geolocation. To this end, many efforts have been made, while many IP geolocation methods still suffer from unacceptable error distance because of the lack of landmarks. In this paper, we propose a novel IP geolocation system, named XLBoost-Geo, which focuses on enhancing the number and the density of highly reliable landmarks. The main idea is to extract location-indicating clues from web pages and locating the web servers based on the clues. Based on the landmarks, XLBoost-Geo is able to geolocate arbitrary IPs with little error distance. Specifically, we first design an entity extracting method based on a bidirectional LSTM neural network with a self-adaptive loss function (LSTM-Ada) to extract the location-indicating clues on web pages and then generate landmarks based on the clues. Then, by measurements on network latency and topology, we estimate the closest landmark and associate the coordinate of the landmark with the location of the target IP. The results of our experiments clearly validate the effectiveness and efficiency of the extracting method, the precision, number, coverage of the landmarks, and the precision of the IP geolocation. On RIPE Atlas nodes, XLBoost-Geo achieves 2,561m median error distance, which outperforms SLG and IPIP.

% motivation
% idea, method(what we did) 研究过程
    % To this end, ...(idea). Specifically, ...(method)
% results 结果和结论

% dict for extracting
\end{abstract}
 
 % their novelty, technical quality, potential impact, insightfulness, depth, clarity, and reproducibility


%% A "teaser" image appears between the author and affiliation
%% information and the body of the document, and typically spans the
%% page.
% \begin{teaserfigure}
%   \includegraphics[width=\textwidth]{sampleteaser}
%   \caption{Seattle Mariners at Spring Training, 2010.}
%   \Description{Enjoying the baseball game from the third-base
%   seats. Ichiro Suzuki preparing to bat.}
%   \label{fig:teaser}
% \end{teaserfigure}

%%
%% This command processes the author and affiliation and title
%% information and builds the first part of the formatted document.

\let\oldmaketitle\maketitle
\renewcommand{\maketitle}{%
  \oldmaketitle%
  \thispagestyle{fancy}}
\maketitle

% KDD
\lhead{Applied Data Science Track Paper}
\rhead{XXX '20, August 22-27, 2020, San Diego, CA, USA}

\section{Introduction}
For the past decades, with network security incidents occur frequently, people have increasingly emphasized on cyberspace security. Due to the critical importance of high-precision IP geolocation in the field of cyberspace security, maintaining cyberspace security has become a main driving force for people to conduct this research. 

%Except for cyberspace security, IP geolocation plays an essential role in many other fields, including Internet finance, Internet advertising, social networking, network performance optimization, location-based content services, etc.

IP geolocation refers to locating the geographical position of Internet devices, % including PC, smartphone, web server, web camera, router, network printer, etc.
most of which do not have a geolocation function themselves. Even for devices with GPS modules, to protect privacy or cover up illegal behavior, they would not share geographical position. Hence, locating Internet devices through IP address becomes one of the most important approaches to discover their geographical position.

% problem and challenge
However, a critical challenge for IP geolocation is how to improve the number of landmarks, Internet devices with a known geographical position. Although existing IP geolocation methods have made great progress, there is still much room for further improvement in positioning precision. Many IP geolocation approaches with high precision rely heavily on the density of landmarks. For example, network measuring based methods, like street-level geolocation (SLG\cite{slg}), map a target device to the nearest landmark's location. Data mining based methods, like Structon\cite{structon} and Checkin-Geo\cite{checkin_geo}, mine landmarks and expand the coverage by clustering IPs. Existing methods that claim to have achieved thousands or hundreds of meters medium error distance (MED), still influenced by regional imbalance and provide unreliable results in regions with sparse landmarks. Therefore, enhancing the number of landmarks is the key to further improve the performance of IP geolocation system.

\begin{figure}[htb]
    \centering
    \includegraphics[scale=0.35]{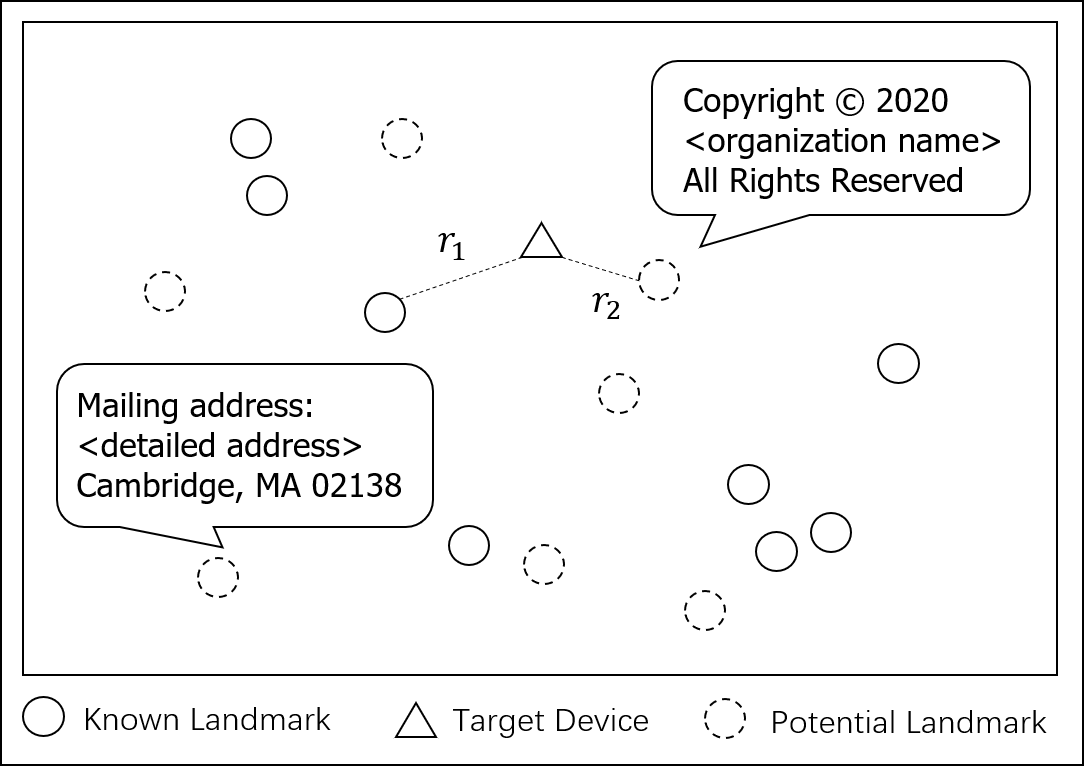}
    \caption{A motivating example: to find potential landmarks as more as possible, by which we can improve the estimation precision}
    \label{fig:motivating_example}
\vspace{-0.1in}
\end{figure}

% 研究内容
To this end, in this paper, we design a novel IP geolocation system, named XLBoost-Geo, focusing on enhancing the number and the density of highly reliable landmarks, which we refer to as landmark boosting. Figure~\ref{fig:motivating_example} shows a motivating example of our IP geolocation system. As we can see, through network measuring, we can find the closest landmark to the target device and map its location to the target as an estimated position. If the bold circles are all landmarks that we can get by previous methods, we can get a position with error distance as $r_1$. However, there are still many potential landmarks (shown as dashed circles) left, which can be mined to further decrease the estimation error of IP geolocation. Therefore, an ideal scenario of XLBoost-Geo is that it can find potential landmarks as more as possible so that we can map the target device to a closer landmark's location. In this case, we can lower the error to $r_2$ as shown in Figure~\ref{fig:motivating_example}. 

In order to mine a large volume of landmarks in a cost-effective way, XLBoost-Geo extracts geographical location clues from web pages on locally deployed web servers. Due to a strong association between the position of local web servers and location clues on their web pages, many previous IP geolocation methods (e.g. SLG\cite{slg} and Structon\cite{structon}) mine landmarks based on this association, but they do not fully utilize this kind of open resources. SLG uses keywords and zipcodes to search for landmarks by mapping services. Even though this method can generate a set of landmarks with a highly precise location, the number is limited because it is hard to cover all possible keywords and maps only include the main portal website for each organization, which is usually hosted on cloud servers and thus can not be used as a landmark. Structon mines geolocation information from 500 million pages, which is large enough to generate plenty of landmarks. However, it uses regular expressions for geolocation extraction, which is able to extract coarse-grained geolocation items, such as city, province, and state, but not street-level and building-level ones. Because of defects during their mining process, the number of landmarks SLG mined is limited, and the location of landmarks mined by Structon is very coarse and can only achieve city-level precision. 

Different from the aforementioned two approaches, XLBoost-Geo scans all addresses in IPv4 space to crawl web pages that contain location-indicating clues. XLBoost-Geo extracts the clues as precise as possible by a variant of the state-of-the-art entity extraction model. For web servers with explicit location information (e.g. contact address) on the web pages, XLBoost-Geo extracts and geocode the complete address to geographical coordinates directly. For those without contact addresses, XLBoost-Geo utilizes implicit clues (e.g. owner name extracted from copyright, title, and logo), which are ignored by previous works, to infer the coordinate. Therefore, XLBoost-Geo is able to boost the number of highly reliable landmarks by extremely making use of clues on web pages.

\begin{figure*}[htb]
    \centering
    \includegraphics[scale=0.48]{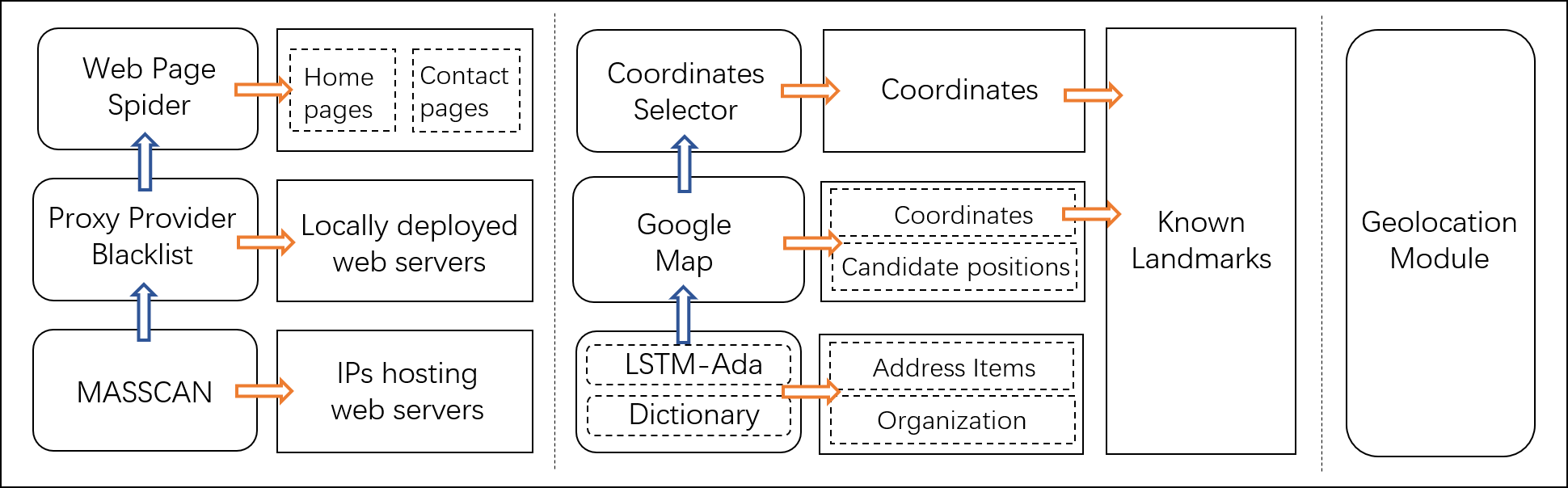}
    \caption{The framework of XLBoost-Geo}
    \label{fig:overview}
\end{figure*}

Figure~\ref{fig:overview} shows the framework of XLBoost-Geo. Specifically, given an IP segment, we first scan and detect IPs that expose HTTP/HTTPS relevant ports by MASSCAN\cite{masscan}. Then, we extract these IPs' organization name from Whois register information and filter locally deployed web servers by a blacklist of proxy providers. Next, we crawl the home pages and contact pages of the local web servers as raw resources for extracting location-indicating clues. Inspired by LSTM-LSTM-Bias\cite{zheng2017joint}, the state-of-the-art model for joint extraction, we design a novel location-indicating clue extracting method based on bidirectional LSTM neural network with a self-adaptive loss function (LSTM-Ada). To supplement the owner names that the LSTM-Ada model fails to extract, we use an organization dictionary (refer to Appendix~\ref{sec:org_dict}) to help. With these clues, we can yield a landmark database by mapping services and further expand the landmark database by a measurement-based coordinate selection algorithm. Based on these landmarks, XLBoost-Geo can locate a target IP by associate the target with the nearest landmark's position. 

Finally, we conduct extensive experiments on real-world IP geolocation tasks. The results clearly validate the effectiveness and the precision of XLBoost-Geo, which has been deployed in a real-world commercial IP geolocation system, named IPPLUS360, which was originally developed based on SLG\cite{slg}. XLBoost-Geo receives remarkable performances in terms of efficiency of landmark mining and enhances the locating precision of IPPLUS360 to a large degree.

% key idea，分步阐述应用的技术和原理，给出系统的overview（架构图）

% 实际部署的效果

% good sentences to imitate
\section{XLBoost-Geo}
% In this section, we introduce three main components of our XLBoost-Geo system in detail: 1) \textit{Data Mining Based Landmark Generation}, \textit{Coordinate Selecting Algorithm Based Landmark Generation} and \textit{IP geolocation}.

% 基于数据挖掘的地标生成
\subsection{Data Mining Based Landmark Generation}\label{sec:landmark_mining}
To construct our landmark database, we first scan a batch of IPs and find out Web servers by detecting HTTP/HTTPS relevant ports. Second, we query the organizations of the IPs by Whois database. Then, by a proxy provider blacklist, we filter out proxy servers which are not hosted locally. The blacklist was generated by manual collection and a classifier trained by Wikipedia description. Since location-indicating clues are only exposed on home pages and contact pages, we only crawl these pages from the local Web servers as raw data. Third, inspired by the state-of-the-art joint extraction model, namely LSTM-LSTM-Bias, we design a variant model with self-adaptive loss function (LSTM-Ada) to extract location clues from the web pages. Finally, based on the clues, we deduce the locations of the web servers and map them to their corresponding IP. The first two steps are easy to implement and not the focus of this paper, so this section details the last two steps.

To extract location clues, we model the process as a sequence tagging task. Since the LSTM-based end-to-end model is widely used in this task and has been shown the effectiveness, we use Bi-LSTM as the main structure of both the encoding layer and the decoding layer. Different from previous models, we employ a self-adaptive loss function that is able to adjust the weights of tags in each training epoch. The self-adaptive loss function has the model more focus on tough-to-extract clues and thus it can enhance the performance of extracting address information on the pages, especially the detailed address (street-level or building-level location-indicating clues). The detailed address is harder to extract than the other location-indicating clues due to its diverse patterns. 

\begin{figure}[htb]
    \centering
    \includegraphics[scale=0.4]{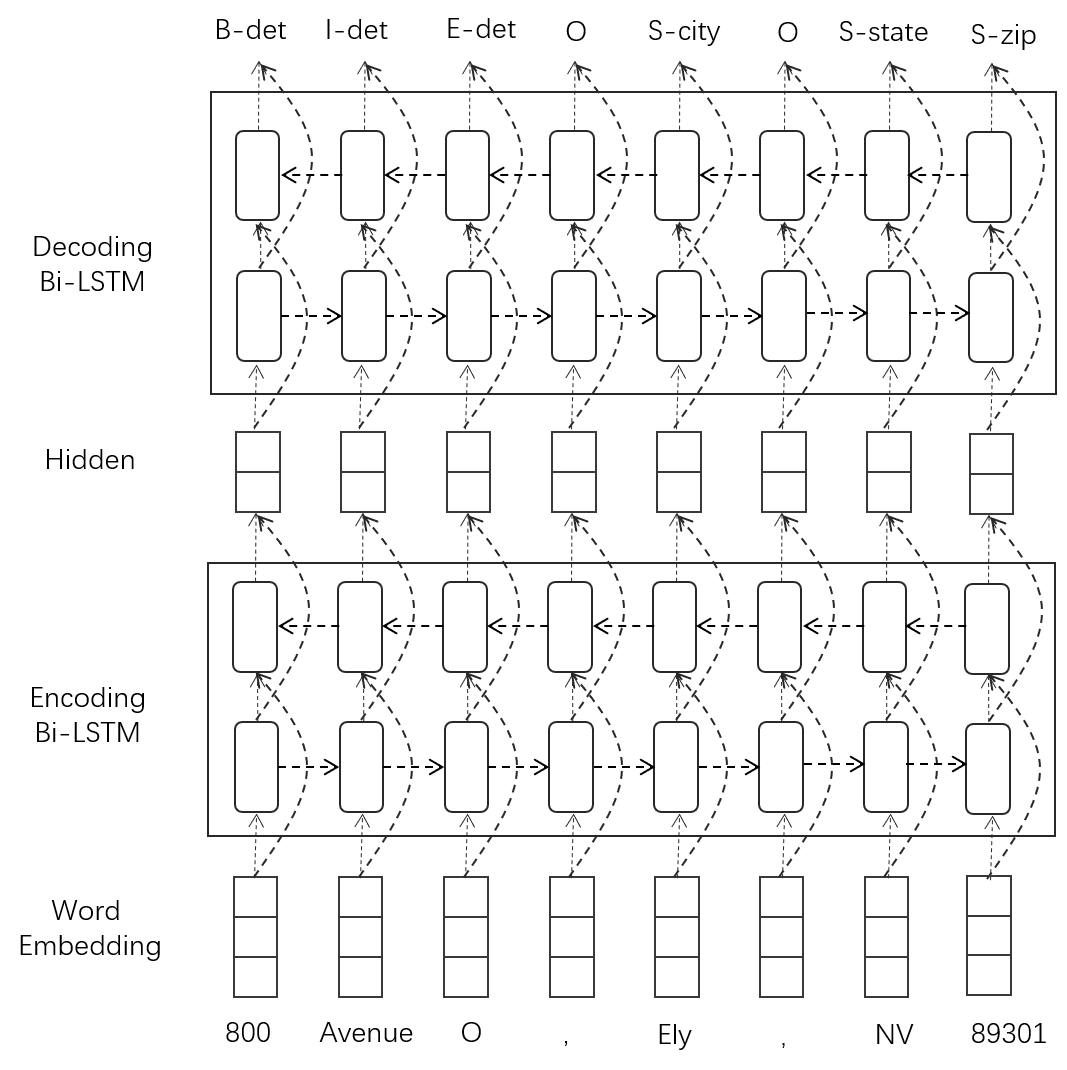}
    \caption{The structure diagram of LSTM-Ada}
    \label{fig:lstm_lstm_ada}
\vspace{-0.1in}
\end{figure}

% encoding layer
The structure diagram of the model is shown in Figure~\ref{fig:lstm_lstm_ada}. The inputs are one-hot vectors of words and the outputs are predicted tags that contribute to extract the location-indicating entities in the page text. The word embedding layer converts the one-hot vector to an embedding vector. Hence, by the word embedding 
layer, a text can be represented as $S = {x_1, ..., x_t, ..., x_n}$, where $x_t$ is the embedding vector corresponding to 
the t-th token in the sentence. After the word embedding layer, there is a Bi-LSTM encoding layer, which consists of two parallel LSTM layers: forward LSTM layer and backward LSTM layer. The LSTM architecture consists of a set of recurrently connected memory blocks, referred to as cells.
The cell in Bi-LSTM encoding layer is used to compute current hidden vector $h_t$ based on the previous hidden vector 
$h_{t-1}$, the previous cell vector $c_{t-1}$ and the current input embedding vector $x_t$. The detail operations for updating hidden state $h_t$ are defined as follows:

\begin{equation}
    \label{equ:input_gate}
    i_t = \sigma(W_i h_{t-1} + U_i x_t + b_i)
\end{equation}

\begin{equation}
    \label{equ:forget_gate}
    f_t = \sigma(W_f h_{t-1} + U_f x_t + b_f)
\end{equation}

\begin{equation}
    \label{equ:cell_hat}
    \Tilde{c_t} = \tanh{(W_c h_{t-1} + U_c x_t + b_c)}
\end{equation}
    
\begin{equation}
    \label{equ:cell}
    c_t = f_t\odot c_t-1 + i_t \odot \Tilde{c_t}
\end{equation}

\begin{equation}
    \label{equ:out_gate}
    o_t = \sigma(W_o h_{t-1} + U_o x_t + b_o)
\end{equation}

\begin{equation}
    \label{equ:hidden}
    h_t = o_t \odot \tanh{(c_t)}
\end{equation}

Here, $\sigma$ is the element-wise sigmoid function and $\odot$ is the element-wise product. $U_*$ and $W_*$ denote the weight matrices and $b_*$ are the bias terms. $x_t$ is the word embedding vector at time $t$, and $h_t$ is the hidden state vector storing contextual information. For each word embedding vector $x_t$, the forward LSTM layer will encode $x_t$ to $\overrightarrow{h_t}$ by considering the contextual information from $x_1$ to $x_t$. The backward LSTM applys the same logic backward and yield $\overleftarrow{h_t}$, which contains contextual information from $x_n$ to $x_t$. Finally, we concatenate $\overrightarrow{h_t}$ and $\overleftarrow{h_t}$ to represent word t’s encoding information, denoted as the bidirectional hidden state $\hat{h_t}$ = [$\overrightarrow{h_t}$; $\overleftarrow{h_t}$].

% decoding layer
We also employ a Bi-LSTM structure for the decoding layer. All the hidden states $\hat{h_t}$ obtained from the Bi-LSTM encoding layer are fed into the Bi-LSTM decoding layer to produce the tag sequence. The operations for calculating the hidden state $h_t$ in the decoding layer are the same as the one in the encoding layer. Based on the $\hat{h_t}$ obtained from the Bi-LSTM decoding layer, the final fully connected layer (FCL) with softmax function computes normalized entity tag probabilities as follows:

\begin{equation}
    y_t = W_y \hat{h_t} + b_y
\end{equation}

\begin{equation}
    p_{t, i} = \frac{e^{y_{t, i}}}{\sum_{j=1}^{N} e^{y_{t, j}}}
\end{equation}

Above, $W_y$ is the weight matrix and $b_y$ is the bias term. $N$ is the total number of tags. $y_{t, j}$ is the j-th element of $y_t$, which is an N-dimension vector output by the FCL. $p_{t, i}$ is the score of assigning the t-th word with the i-th tag. 

% self-adaptive loss function
During training, we use Adam\cite{kingma2014adam} to maximize the log-probability of the data. The objective function is defined below:

% 目标函数公式
\begin{equation}
    \label{equ:loss}
    L = - \sum_{s=1}^{|D|} \sum_{t=1}^{T} w \odot \hat{y}_t^s \sum_{j=1}^{N} \hat{y}_{t,j}^s log(p_{t,j}^s)
\end{equation}

\begin{equation}
    \label{equ:weight}
    w_i = exp(\alpha(\frac{1}{N} \sum_{j=1}^N {F_1}^j - {F_1}^i))
\end{equation}

% 解释目标函数中的符号
% 解释权重
Here, $|D|$ is the size of a batch of data, $T$ is the length of a page text, $N$ is the number of tags, $\hat{y}_t^s$ is the one-hot vector of the true label of the t-th word in the s-th text, and $\hat{y}_{t,j}$ is the j-th element of $\hat{y}_t^s$. Besides, $w_i$ is the i-th element of $w$, which is the weight vector of tags, ${F_1}^i$ is the $F_1$ score of the i-th tag on the validation set in the last epoch, and $\alpha$ is the coefficient of distinguishing, which adjusts the degree of distinguishing the weights of tags.

% tagging scheme
    % BIOES
    % 举例
During the inference, we label each word as the tag with the highest score. For each page text, we obtain a tag sequence from the model, based on which we can extract location-indicating entities from the page text. We use the “BIESO” (Begin, Inside, End, Single, Other) signs to represent a word's position in the entity. The tag "O" means that the corresponding word has nothing to do with the location-indicating entity. Different from “O”, the other tags consist of two parts: the word position in the entity and the entity type, which are concatenated using a dash. As shown in Figure~\ref{fig:lstm_lstm_ada}, the text "800 Avenue O, Ely, NV 89301" is fed into the model to output the tag sequence {B-det, I-det, E-det, O, S-city, O, S-state, S-zip}, which means "800 Avenue O" is the detailed address, "Ely" is a city, "NV" is a state, and 89301 is a ZIP code.
% 例子待补全，根据新画的模型图

% word embedding: fasttext 原因 -> 放到实验参数里说

% ——————————————————————————————————————————————————————————————————————————————
% how to use clues
    % complete addresses,  (detail, city, state, ZIP code)
    % organization names and incomplete address
        % google map, search org name in the incomplete address(zipcode, state, city)
    % organization name only (website without revealing address information)
        % ping measurement by RIPE Atlas probes
        % several probes to estimate a consecutive region (CBG)
          % google map, if only one candidate coordinate -> landmar
          % more than one candidate coordinate -> IP Geolocation

% 可以提取出5种clues
Now given a web page with contact information and copyright information, we can extract at most 5 kinds of location-indicating items, including organization name, ZIP code, state, city, street-level or building-level address (detailed address for short). Figure~\ref{fig:motivating_example} also exemplifies these location-indicating items. If a web page contains these 5 items, we can extract all of them in most cases with LSTM-Ada. Then, we can order them into a formatted address and geocode it to a geographical coordinate by a mapping service(e.g. Google Map). In fact, full information is not a must. Four-Tuple (detailed address, city, state, ZIP code) is enough to decide a coordinate. In other words, if a page contains full contact information, the organization name is not necessary. 

% 如果提取不到detailed address，用organization name与区域线索也可以定位
However, Our experiment results show that the detailed address is the bottle-neck of the extraction process but the organization name is more common and easier to extract. Plus, an organization usually has only one or several subsidiaries in a small region. Therefore, if a web page only exposes the organization name but not contact information, or if LSTM-Ada fails to extract the full contact information, especially the detailed address, the organization name will play its magic in the geolocation process. If any clues indicating a small region, e.g. state, city, or ZIP code, can be successfully extracted, we can, through a map, look up several possible coordinates in the region with the organization name as a keyword. If the model fails to extract any region-indicating clues, we use a variant of CBG\cite{cbg} (refer to Appendix~\ref{sec:cbg_var}) to help determine possible coordinates. If there is only one possible coordinate returned, we map it to the IP directly. If there is more than one possible coordinate returned, we leave the rest tasks to the coordinate selection algorithm which is mentioned in Section~\ref{sec:selection_algorithm}.

% 基于坐标选择算法的地标生成
\subsection{Coordinate Selecting Algorithm Based Landmark Generation}\label{sec:selection_algorithm}
% 算法基于两个假设
If a possible region is determined, and there is more than one possible coordinate in the region, we will select the most possible one based on measurements. This algorithm is based on two assumptions. First, The closer the two Internet devices are the closer the delays from the same probe to them. Second, The closer the two Internet devices are, the shorter the route from one to another. Even though these two assumptions are not always true, they work very well in previous IP geolocation methods, e.g. CBG, SLG, and etc. 

% 算法的大体步骤
This algorithm mainly includes four steps. First, in the vicinity of possible coordinates, select a set of landmarks and a set of probes for estimating the distance between each landmark and the target IP. Second, score the landmarks based on the estimated distance to the target IP. Third, deliver the score from landmarks to the candidate coordinates according to the geographical distance. Finally, select the location with the highest score and map it to the target IP. Based on the two assumptions mentioned above, we estimate the distance between each landmark and the target IP by calculating the similarity between the delay vectors and computing the length of the shortest route. As for the geographical distance between a landmark and each candidate coordinate, we obtain it by calculating the great circle distance \cite{vincenty1975direct} between the two coordinates.

% 候选地标得分的具体计算过程和公式
Formally, the score of a candidate coordinate can be calculated as below:
\begin{equation}
\label{equ:final_score}
score({c_i}) = \sum\nolimits_{{l_j} \in L}{g({l_j, c_i})}{s({l_j})} 
\end{equation}
Here, $s(l_j)$is the score of each landmark$l_j$ and $g(l_j,c_i)$is the redistributing gate, which play the role of delivering a weighted $s(l_j)$ to every candidate coordinate $c_i$. The weight is based on the distance between the landmark and the candidate coordinate.

As mentioned before, a candidate should get a higher score from a landmark if it is closer to the landmark than the others. Therefore, the redistributing gates are defined below:
\begin{equation}
\label{equ:redistribution_gate}
g({l_j, c_i}) = 1 - \frac{e^{dis({c_i},{l_j})}}{{\sum\nolimits_{{l_k} \in L} {{e^{dis({c_i},{l_k})}}}}} 
\end{equation}

Here, $c_i$ represents a candidate coordinate, $l_i$ represents a landmark, and $dis(c_i, l_i)$ refers to the great circle distance between them. 

Based on the two previously mentioned assumptions, we use the following formula to calculate the score of each landmark:
\begin{equation}
    \label{equ:scr_toge}
    s({l_i}) = \alpha \cdot{s_d}({l_i}) + \beta \cdot{s_t}({l_i})
\end{equation}

Above, ${s_t}({l_i})$ and ${s_d}({l_i})$ represent the score of a landmark in the perspective of topology and the perspective of pure network delay measurement. The former is decided by the length of the shortest route from the landmark to the target IP. The latter is decided by the similarity between the delay vector of the landmark and the delay vector of the target IP. $\alpha$ and $\beta$ are the weights of ${s_d}({l_i})$ and ${s_t}({l_i})$ separately.  

The score of a landmark in the perspective of pure network delay measurement can be calculated as following two formulas:
\begin{equation}
    \label{equ:scr_de}
    {s_d}({l_i}) = \frac{sim({l_i},t)}{\sum\nolimits_{{l_j} \in L}{sim({l_j},t)} }
\end{equation}

\begin{equation}
    \label{equ:cos_sim}
    sim({l_i},t) = \frac{{\boldsymbol{v}_{l_i}}{\boldsymbol{v}_{t}}}{\left|{\boldsymbol{v}_{l_i}}\right| \left|{\boldsymbol{v}_{t}}\right|}
\end{equation}
    
${v}_{l_i}$ and ${v}_{t}$ denote the delay vector of a landmark and the target IP separately. Each element in the vectors is the delay measured from a probe. ${sim({l_i}, t)}$ represents the cosine similarity between ${v}_{l_i}$ and ${v}_{t}$.

\begin{figure}[htb]
    \centering
    \includegraphics[scale=0.36]{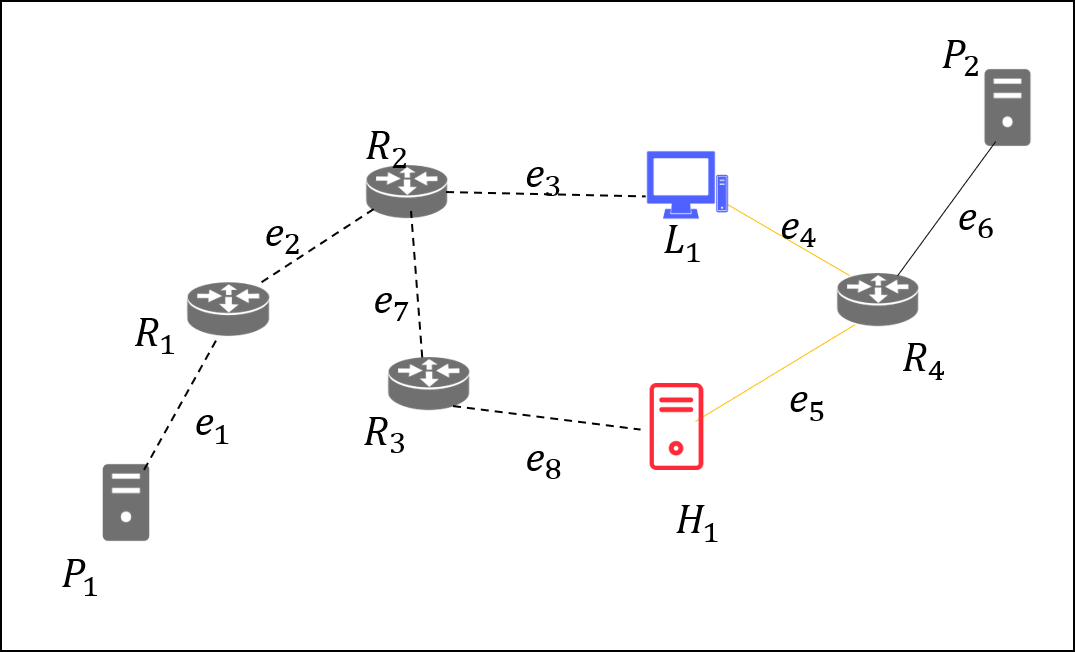}
    \caption{To approximately calculate the length of the shortest route by finding the closest router}
    \label{fig:shortest_route}
\vspace{-0.1in}
\end{figure}

The score of a landmark in the perspective of topology can be calculated as below:
\begin{equation}
    \label{equ:scr_topo}
    {s_t}({l_i}) = 1 - \frac{{e^{r({l_i},t)}}}{\sum\nolimits_{{l_j} \in L} {{e^{r({l_j},t)}}} }
\end{equation}
$r({l_i},t)$ denotes the length of the shortest route between a landmark $l_i$ and the target IP $t$. Inspired by \cite{slg}, we approximately calculate the shortest route by finding the closest common router. Figure~\ref{fig:shortest_route} illustrates an example, we first send traceroute probes to the landmark $L_1$ and the target host $H_1$ from all probes ($P_1$, $P_2$, and etc.). For each probe, we then find the closest common router to $H_1$ and $L_1$, shown as R2 and R4 in Figure~\ref{fig:shortest_route}. Through the common routers (R2 and R4), The routes passing from $L_1$ to $H_1$ is the approximate indirect route. In Figure~\ref{fig:shortest_route}, The indirect route $r_1$ is $L_1$ -> $R_2$ -> $R_3$ -> $H_1$, and the indirect route $r_2$ is $L_1$ -> $R_4$ -> $H_1$. We separately sum the delays on the routes as the their length. Since $e_4$ + $e_5$ is shorter than $e_3 + e_7 + e_8$, we take the former as the length of the shortest route.

\subsection{IP Geolocation Based on Landmarks}
Based on the landmarks mined by the aforementioned method, XLBoost-Geo is able to geolocate an arbitrary IP. The main idea is to find the closest landmark to the target IP and take the location of the landmark as an estimate of the target IP. First, select a set of probes that closest to the target IP based on the assumption that the smaller the delay, the closer the probe to the target. Second, calculate the scores of all landmarks according to Formula~\ref{equ:scr_de} and select a set of candidate landmarks with the highest scores. Third, calculate the scores of the candidate landmarks according to Formula~\ref{equ:scr_topo} and sum two scores together as Formula~\ref{equ:scr_toge}. Then, select the landmark with the highest score as the closest landmark.

\section{Experimental Result}
\subsection{The Performance of Location-indicating Clue Extraction}\label{sec: exp_clue_extraction}
\textbf{Data Description:} As previously mentioned, we use LSTM-Ada to extract the location-indicating clues on web pages. To evaluate the performance of our model, we produce a web page dataset, namely WPLICE (Web Pages for Location-indicating Clues Extraction), which consists of 269,566 web pages that contain location-indicating clues. These pages are crawled from yellow pages websites and web mapping services, which record plenty of organizations with their website, address, and name. To prepare this dataset for training, validating, and evaluation, we extract texts from the web pages and semi-automatically labeled the location-indicating entities on the texts. When it comes to "semi-automatically", we mean automatically labeling the page texts based on the organization name and address information revealed and then revising the labels manually. We split this dataset into a train set and a test set by ratio 7:3 and randomly sample 10\% data from the test set for validating and choosing the best model.  

\textbf{Experimental Setup:} In the experiment, we adopt standard Precision (Prec), Recall (Rec) and F1 score to evaluate the performance of our model. To highlight the strengths of our model in the location-indicating entity extraction task, we compare its performance with two popular sequence tagging models, LSTM-CRF and LSTM-CNN. To highlight the key role of the self-adaptive loss function, we not only calculate the scores on overall extraction results but also calculate the scores of different kinds of tags.

\textbf{Hyper-parameters:} We use FastText\cite{bojanowski2017enriching} to pre-train the word embeddings on 22,294,645 web page texts and 11,073,587 addresses. The dimension of the embedding vector is set to 50. The dimension of the embedding layer is also set to 50 and initialized by the embeddings trained by the FastText, the number of units of the Bi-LSTM encoding layer is set to 256 and the number of units of the Bi-LSTM decoding layer is set to 512. The distinguishing coefficient $\alpha$ corresponding to the results in Table~\ref{table:entity_extraction} is 64. The model is first optimized by using Adam\cite{kingma2014adam} and then fine-tuned by using stochastic gradient descent (SDG). Refer to Appendix~\ref{sec:hyper_para_model} for more details about setting hyper-parameters.

\begin{table}[htbp]
\renewcommand{\arraystretch}{1.1}
\caption{The performance of location-indicating entity extraction on WPLICE}
\label{table:entity_extraction}
\centering
\begin{tabular}{|c|c|c|c|c|c|}
\hline
 Models & Results & Prec & Rec & F1 & Acc\\
\hline
\hline
         & all types & 0.9757 & 0.9715 & 0.9736 & - \\
         & organization & 0.9771 & 0.9683 & 0.9727 & 0.9632\\
         & detailed &  0.9728 & \textbf{0.9536} & \textbf{0.9631} & \textbf{0.9416}\\
LSTM-CRF & city & 0.9721 & 0.9666 & 0.9693 & 0.9641\\
         & state & 0.9811 & 0.9827 & 0.9819 & 0.9867\\
         & ZIP code & 0.9799 & 0.9795 & 0.9797 & 0.9734\\
         & full info & - & - & - & \textbf{0.8882}\\
\hline    
         & all types & 0.9764 & 0.0.9688 & 0.9725 & -\\
         & organization & 0.9780 & 0.9662 & 0.9720 & 0.9698\\
         & detailed & 0.9731 & \textbf{0.9546} & \textbf{0.9637} & \textbf{0.9422}\\
LSTM-CNN & city & 0.9716 & 0.9625 & 0.9670 & 0.9628\\
         & state & 0.9829 & 0.9789 & 0.9809 & 0.9842\\
         & ZIP code & 0.9775 & 0.9756 & 0.9765 & 0.9737\\
         & full info & - & - & - & 0.8792\\
\hline    
         & all types & 0.9690 & 0.9812 & 0.9750 & -\\
         & organization & 0.9689 & 0.9798 & 0.9743 & 0.9675\\
         & detailed & 0.9682 & \textbf{0.9752} & \textbf{0.9716} & \textbf{0.9661} \\
LSTM-Ada & city & 0.9692 & 0.9775 & 0.9732 & 0.9682\\
(our model) & state & 0.9694 & 0.9942 & 0.9816 & 0.9874\\
         & ZIP code & 0.9724 & 0.9822 & 0.9772 & 0.9786\\
         & full info & - & - & - & \textbf{0.8937}\\
\hline
\end{tabular}
\end{table}

% 训练数据集较大时，CRF的优势并不明显, LSTM decoding layer 也可以很好地学习到各个标签之间的转移关系
% 因为使用的是fasttext训练的词向量，考虑了字词，与CNN的效果类似，所以LSTM-CNN的优势也不明显
% 其他两个模型虽然在个别标签上有较高的得分，但得分的不均衡性导致whole info受到限制。
% 从表格看，self-adaptive loss function 使得 Rec显著增高，Prec略微下降。这是因为标签“O”总是获得相对较低的权重，使得模型更大胆地标注出实体。

\textbf{Results and analysis:} The performances on location-indicating entity extraction are shown in Table \ref{table:entity_extraction}. In the table, \textit{Acc} denotes the accuracy at page-level, which can be calculated by Formula~\ref{equ:acc_p}, where $P_suc$ denotes the number of web pages of which the clue is successfully extracted, and $P_total$ denotes the number of total pages. Besides, "all types" means to calculate scores without distinguishing each entity type, and "full info" means all 5 types of entities need to be extracted from the page. Clearly, we observe that our model outperforms the other two models on "detailed address" and "full info", which demonstrates the effectiveness of the self-adaptive loss function. As the function helps to focus on tough-to-extract entities, especially the entities of the detailed address, it help the model improve the success rate of extracting the full location-indicating information. Besides, we use FastText to train word embeddings, which also take into account character-level features, thus the LSTM-CNN has little competitive edges. The dataset is huge enough for the LSTM to learn the transition rules of the tags, so the CRF layer provides little advantages to the model LSTM-CRF. Since the more location-indicating information we get the more precise the location we can get, we use LSTM-Ada to extract location-indicating clues.

\begin{equation}
    \label{equ:acc_p}
    Acc = \frac{P_suc}{P_total}
\end{equation}

\subsection{The Performance of Landmark Mining}
\textbf{Data Description:} 
% Planet Lab nodes 介绍，为什么选择这些数据（web servers with known locations）
% 数据怎么做的（详细：IP映射、过滤美国、为什么美国）
For evaluating the accuracy of the locating of our landmark mining method, we collected data about web servers on PlanetLab\footnote{https://www.planet-lab.org/db/pub/sites.php} to generate the ground-truth dataset. Since the nodes on this Platform report their websites and geographic coordinates, we compare the coordinate estimated by our method and the one shown on PlanetLab. We map the websites to IPs by DNS and remove nodes not locally deployed by the proxy blacklist. Since cross-language problems on entity recognition are not what this paper's focus, we select the nodes located in the United States to only deal with English websites for convenience. Finally, we got 104 nodes left. We referred to this dataset as PlanetLab-104. 

% http/https web banners（contact pages/home pages, 过滤美国，原因同上）
To show the number, coverage, and density of the landmarks that XLBoost-Geo is able to mine, we scaned 1.5 billion IPs of the US by MASSCAN\footnote{https://github.com/robertdavidgraham/masscan} and discovered 6,716,338 local web severs. We crawled home pages and contact pages from these nodes and totally got 8,723,627 pages as the raw data for mining. We referred to this dataset as WebPages-8M. 

\textbf{Experiment Setup:} 
% 实验：要做怎样的实验，目的是什么，指标是什么
    % MED：定位误差
    % 数量和密度
 % 对照：
    % 对于planet nodes，对照的baseline：IPIP、SLG的商业工具，为什么选这两个
    % 对于banners，挖出的地标对比SLG通过zipcode和keyword找到的数量   
In this section, we show the results of two experiments. For PlanetLab-104, we adopt median error distance (MED), which is commonly used in previous work, to evaluate the accuracy of the estimated location. For comparison, we select a commercial tool IPIP and the state-of-the-art IP geolocation method SLG as baselines. Besides, we mine landmarks from the WebPages-8M and compare the number of landmarks with the ones mined by SLG's landmark mining method, which is based on keywords and ZIP code.
% 参数设置：probes和landmarks的数量，score的两个参数：alpha、beta（相关描述和特殊情况处理 写到这个部分）
In the experiments, we set to 100 the number of probes in the measurements for the CBG variant(Section~\ref{sec:landmark_mining}). In coordinate selection algorithm(Section~\ref{sec:selection_algorithm}), We set to 200 the number of probes and set to 1000 the maximum number of the landmarks that closest to the candidate coordinates. Since some intermediate hops do not respond in some cases, the probes can not detect the complete routes to the target and landmarks. In the Formula~\ref{equ:scr_toge}, if ${s_t}({l_i})$ is not available owing to the incomplete route, $\alpha$ and $\beta$ are set to 1 and 0 separately. In normal cases, both of them are set to 0.5 in XLBoost-Geo.
    
\begin{figure}[htb]
    \centering
    \includegraphics[scale=0.5]{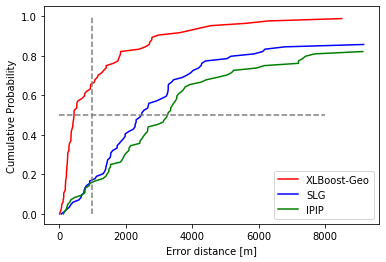}
    \caption{The cumulative distribution function of the error distance on PlanetLab-104}
    \label{fig:ed_planetlab}
\vspace{-0.1in}
\end{figure}

\begin{figure}[htb]
    \centering
    \includegraphics[scale=0.28]{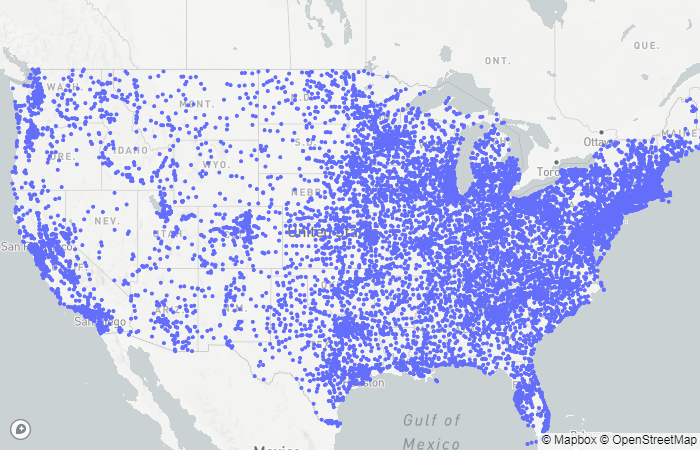}
    \caption{The distribution of the landmarks in the mainland of the US}
    \label{fig:landmarks_distribution_us_110w}
\vspace{-0.1in}
\end{figure}

\textbf{Results and analysis:}
% Planet nodes
The performance of the accuracy is shown in Figure~\ref{fig:ed_planetlab}.
As the figure shows, XLBoost-Geo outperforms SLG and IPIP on the estimation accuracy. Specifically, the MED of XLBoost-Geo, SLG, IPIP are 441m, 2,446m, 3,242m respectively (the grey horizontal dashed line). In the results provided by XLBoost-Geo, the percentage of nodes with less than 1km error distance is more than 60, which is more than 3 times as the other two methods (the grey vertical dashed line).
% SLG uses nearby landmarks to locate this target nodes, but XLBoost-Geo can locate the real location of these targets and turn them into landmarks.
% 结论：XLBoost-Geo 可以挖掘的地标具有较高精度
The above results demonstrate that XLBoost-Geo has the ability to mine highly-reliable landmarks with little location error.
% mine landmarks from web banners
% 结论：XLBoost-Geo 可以尽最大努力挖掘海量地标并且覆盖范围广泛、密度较高
We totally got 1,115,076 landmarks by XLBoost-Geo from the WebPages-8M. Figure~\ref{fig:landmarks_distribution_us_110w} shows the distribution of the landmarks in the mainland. As shown by the Figure, except for the Midwest where there are few inhabitants, the eastern and western coastal landmarks are very dense. Besides, The landmarks cover 87\% counties in the mainland. The above results show the ability of XLBoost-Geo to mine a large number of and widespread landmarks. Besides, Group by 10 randomly selected ZIP codes, Table~\ref{table:comparison_landmarks_zipcodes} shows the comparison of the number of landmarks between the landmark mining method of XLBoost-Geo and SLG's landmark mining method. In the table, $X$ and $S$ are the abbreviations of XLBoost-Geo and SLG separately, and POI represents "Point of Interests" searched by map services. After collecting POIs with keywords, SLG's method has to remove non-locally deployed nodes from the POIs to filter valid landmarks. As the result shows, SLG's method can only mine tens of landmarks per ZIP code. In contrast, XLBoost-Geo can mine hundreds. The intuitive reason is that SLG's method is limited by the keywords they use and the number of websites on the web map. This result shows the distinguished efficiency of XLBoost-Geo on landmark mining. 

% \begin{figure}[htb]
%     \centering
%     \includegraphics[scale=0.35]{Figures/landmarks_distribution_county.png}
%     \caption{Approximately calculate the shortest route between a landmark and the target IP by finding the closest router}
%     \label{fig:landmarks_distribution_county}
% \vspace{-0.1in}
% \end{figure}

\begin{table}[htbp]
\renewcommand{\arraystretch}{1.3}
\caption{The comparison of the number of landmarks grouped by ZIP codes.}
\label{table:comparison_landmarks_zipcodes}
\centering
\begin{tabular}{|c|c|c|c|}
\hline
 ZIP code & Landmarks-X & POI & Landmarks-S \\
\hline
\hline
60007 & 210 & 20 & 6 \\
\hline
10018 & 142 & 9 & 1 \\
\hline
10017 & 129 & 28 & 8 \\ 
\hline
10036 & 126 & 16 & 0 \\
\hline
10022 & 125 & 27 & 5 \\
\hline
91730 & 115 & 20 & 4 \\
\hline
20005 & 111 & 17 & 4 \\
\hline
20036 & 111 & 21 & 4 \\
\hline
10001 & 108 & 26 & 3 \\
\hline
10016 & 100 & 24 & 4 \\
\hline
\end{tabular}
\end{table}

\subsection{The Performance of IP Geolocation}
% Ripe Atlas nodes
% 参考 PlanetLab nodes
\begin{figure}[htb]
    \centering
    \includegraphics[scale=0.5]{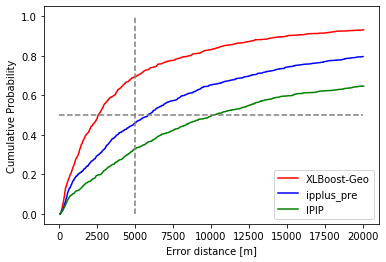}
    \caption{The cumulative distribution function of the error distance on RIPE-Atlas-1156}
    \label{fig:ed_ripe}
\vspace{-0.1in}
\end{figure}
% 因为nodes的网络环境复杂多样
% On this data set, some previous algorithms and current popular IP geolocation tools can not perform well as they reported. In this section, we will compare the performance of XLBoost-Geo with the performance of SLG and IPIP on this dataset.

\textbf{Data Description:}
For evaluating the performance on IP geolocation, we collected the ground-truth dataset from RIPE Atlas\footnote{https://atlas.ripe.net/}. RIPE Atlas is the RIPE NCC’s global network of probes that actively measure Internet connectivity. The Members are required to report the coordinates of the probes they own, so we can easily get a lot of <IP, coordinate> pair from the platform. We select 1156 nodes located in the US to form our dataset, called RIPE-Atlas-1156. This dataset consists of nodes in different network scenes (Table~\ref{table:performance_network_scene}), including residential network (306), corporate private network (414), schools (143), data centers (217), and other organizations (76).

\textbf{Experiment Setup:}
The same as PlanetLab-104, we adopt MED as the metric to evaluate the accuracy of the estimated location. For comparison, we select IPIP and SLG as baselines. 
In the experiments, we set to 200 the number of probes and set to 1000 the number of the candidate landmarks. We deal with $\alpha$ and $\beta$ (Formula~\ref{equ:scr_toge}) the same way as we do in the coordinate selection algorithm.

\textbf{Results and analysis:}
% 描述图片: MED, 5000米以下比例
As shown in Figure~\ref{fig:ed_ripe}, XLBoost-Geo outperforms SLG and IPIP on the estimation accuracy. Specifically, the MED of XLBoost-Geo, SLG, IPIP are 2,561m, 5,828m, 10,121m respectively (the grey horizontal dashed line). As for XLBoost-Geo, SLG, and IPIP, the percentages of nodes with less than 5km error distance are 63, 46, 32 separately (the grey vertical dashed line). The figure demonstrates that even for a dataset consist of nodes in varied network scenes, XLBoost-Geo has the ability to geolocate an IP in high precision. To be more specific, Table~\ref{table:performance_network_scene} shows the performance on nodes in each network scene. In the table, $X$, $S$, and $I$ denotes "XLBoost-Geo", "SLG", and "IPIP" separately. As shown, XLBoost-Geo outperforms SLG and IPIP in all network scenes, especially the data centers and other organizations, which speaks for the important role of the organizations' locally-deployed web servers and also indicates that there are many locally-deployed web servers in the vicinity of data centers. Even though SLG performs as well as XLBoost-Geo and achieves 625 MED on school nodes, due to the regional imbalance of landmarks, it fails to achieve the same precision on PlanetLab-104, which also consists of web servers in Universities. In contrast, XLBoost-Geo achieves similar precision on both PlanetLab-104 and these 143 school nodes. According to the above results, XLBoost-Geo mitigates the negative impact of regional imbalance.

\begin{table}[htbp]
\renewcommand{\arraystretch}{1.3}
\caption{The performance in each network scene.}
\label{table:performance_network_scene}
\centering
\begin{tabular}{|c|c|c|c||c|c|}
\hline
 Scenario & Num & $MED_{X}$ & $MED_{S}$& $MED_{I}$\\
\hline
\hline
residential network & 306 & 3,711 & 5,410 & 11,255 \\
\hline
corporate private network& 414 & \textbf{2,806} & 6,469 & 12,300 \\
\hline
data centers & 217 & \textbf{1,903} & 6,811 & 8,059 \\
\hline
schools & 143 & \textbf{491} & \textbf{625} & 4,179 \\ 
\hline
other organizations & 76 & \textbf{1,402} & 6,670 & 9,067 \\
\hline
total & 1156 & \textbf{2,561} & 5,828 & 10,121 \\
\hline
\end{tabular}
\end{table}

\section{Related Work}
% In this section, We introduce some typical literature and works that relate to our study. They mainly fall into two aspects: \textit{Entity Extraction} and \textit{IP Geolocation}. 

\subsection{Entity Extraction}
In our study, the landmark mining process relies heavily on the accuracy of location clues extracted from web pages. In other words, an efficient and reliable algorithm to extract accurate location clues plays an important role in our study. In our approach, we model the clue mining task to an entity extraction task, which targets at locating location clues in the textual content on web pages. In the past, Entity extraction has been widely studied by researchers. Traditional algorithms, e.g., Hidden Markov Model\cite{zhou2002named} and Conditional Random Fields (CRF)\cite{passos2014lexicon}, usually rely on high-quality hand-crafted features and well-designed models. Recently, with the fast development and popularity of deep learning, several neural network-based sequential token tagging techniques have been proposed to address entity extraction tasks, which requires little manual labor to do feature engineering but achieved better performance. For instance, Lample et al.\cite{lample2016neural} combine bidirectional Long Short-Term Memory networks (Bi-LSTM) and CRF for named entity recognition. Chiu et al. use a hybrid Bi-LSTM and Convolutional Neural Network (CNN) to consider both word-level and character-level features\cite{chiu2016named}.
% 具体工作的一句话概述待替换，看原文摘要或者intro

\subsection{IP Geolocation}
IP geolocation methods can be simply divided into two categories: 1) \textit{Data mining-based Methods} and 2) \textit{Measurement-based Methods}. Data mining based methods are measurement-independent and only rely on mining location clues in source data to locate a target IP. These methods are easy to implement but the precision and the coverage are limited by mining methods and the quality of the source data. Measurement-based methods, which rely on network measurements, are universally applicable to nodes that allow ICMP requests. These methods can cover more IPs than data mining-base methods but suffer from higher computational overhead.

\textbf{Data Mining-based Methods: }
Since regional Internet registry databases(Whois databases) include the address of the organization of a target IP, Moore et al.\cite{whois_geo} query and extract the address as the location of the target IP. Since a domain name can contain different granular descriptions of geographic location information, like the abbreviation of a city, state, or country, by mining this information, the geographic location of the target device can be inferred. Based on this idea, GeoTrack\cite{ip2geo}, DRoP\cite{huffaker2014drop}, rDNS-Geo\cite{dan2018ip_by_dns} and HLOC\cite{scheitle2017hloc} mine location clues in domain names to estimate a location. Structon \cite{structon} uses regular expressions to extract address information from web pages. By mapping addresses to the corresponding IPs of the web servers, it generates hundreds of thousands of landmarks with city-level precision. GeoCluster \cite{ip2geo} uses the address prefixes(AP) in BGP routing tables to cluster IP addresses and then deduce the entire cluster by location information of a few hosts in a cluster. The location information is extracted from three sources, including login and registration information of Hotmail users, HTTP cookies of bCentral, and users' query of FooTV. Checkin-Geo \cite{checkin_geo} leverages the location data shared by users in location-sharing services and logs of user logins from PCs for real-time and accurate geolocation. Dan et al.\cite{ipgeo_query_log} generate a large ground-truth dataset using real-time global positioning data extracted from search engine logs. Using the dataset, they measure the accuracy of three commercial IP geolocation databases. Plus, they introduce a technique to improve existing geolocation databases by mining explicit locations from query logs. Based on a crowd-sourcing idea, Lee et al.\cite{lee2016ip} propose an IP geolocation database creation method utilizing Internet broadband performance measurement tagged with locations, and they present an IP geolocation database based on 7 years of Internet broadband performance data in Korea.

\textbf{Measurement-based Methods: }
GeoPing \cite{ip2geo} maps a node to the nearest probe's location based on the measured delays from probes to the node. CBG \cite{cbg} utilizes a set of probes to measure a set of delays to a target IP. Based on the delays and delay-distance relation, CBG draws a set of constraint circles, of which the overlap region is where the target IP resides. TBG \cite{tbg} claims that routers nearby landmarks are easy to locate and network topology can be effectively leveraged to improve previous IP geolocation techniques. Based on CBG, Octant \cite{octant} takes both positive and negative measurement constraints into account when estimating a target's position. Ciavarrini et al.\cite{ciavarrini2015geolocation}\cite{ciavarrini2017smartphone} devised an active IP geolocation method that uses smartphones as landmarks based on crowd-sourcing principles: a number of users participate voluntarily to the system and provide their devices as measuring devices. Wang et al.\cite{wanggeo} utilize zip codes and keywords to generate passive landmarks through mapping services. They compute the indirect delay between a target and landmarks by finding the closest common routers. Then, they find a landmark with the minimum indirect delay to the target IP and associate the target's location with it. 

Probability and statistics-based methods aim to study the relation between network delay and great circle distance and decrease the error introduced by the delay-distance model. Typical algorithms based on probability and statistics mainly include Posit\cite{eriksson2012posit}, Spotter\cite{laki2011spotter}, GBLC\cite{zhu2016city}, etc. They do not assume that there is a linear relationship between network delay and geographical distance but estimate the statistical relationship between them through statistical analysis of delay-distance data. 

The main idea of Machine learning based methods is to reduce IP geolocation to a machine-learning classification problem. Typical machine learning based methods includes LBG\cite{eriksson2010learning}, ELC\cite{maziku2012enhancing}, CBIG\cite{biswal2014classification}, and NN-Geo\cite{jiang2016ip}. LBG\cite{eriksson2010learning} is a Naive Bayes estimation method that assigns a given IP target to a geographic partition based on a set of lightweight measurements from a set of probes to the target. ELC\cite{maziku2012enhancing} extends LBG by adding three features from measurements and implementing a new landmark selection policy. CBIG\cite{biswal2014classification} employs ELC to improve the accuracy of geolocating data files in datacenters in four commercial cloud providers. With measurement results collected from landmarks, Jiang et al.\cite{jiang2016ip} trained a two-tier neural network that estimated the geolocation of arbitrary IP addresses.

\section{Conclusion}
% 1. 主要发现及其意义/应用；
% 2. 倾向或者推荐是什么；
% 3. 未来的研究方向；
% 4. 研究缺陷。

In this paper, we introduce a novel IP geolocation system, named XLBoost-Geo. The key idea of XLBoost-Geo is to extremely utilize location-indicating clues on web pages to mine a large number of landmarks to help IP geolocation. XLBoost-Geo mainly consists of three components. First, we use LSTM-Ada to extract the location-indicating clues on web pages and generate an initial landmark database based on the fine-grained clues.  Second, to deal with left IPs that have multiple candidate locations, we use the coordinate selection algorithm to select the most possible coordinate. Third, by measurements on network latency and topology, we estimate the closest landmark and associate its coordinate with the target IP. The experiments demonstrate the effectiveness and efficiency of LSTM-Ada on location-indicating clue extracting, the precision, number, coverage of the mined landmarks, and the precision of the IP geolocation. This paper has two contributions: 1) We propose a novel, efficient, and universally applicable landmark mining approach, which utilizes a kind of new information (organization name) previous methods never use. Besides, to make full use of the web pages, we use RNN to extract the clues instead of rule-based methods that previous approaches commonly use. 2) We demonstrate that open data (e.g. the web pages) is enough to help geolocate IPs with very high precision.

% three contributions:
% 提出了一个新颖，高效且通用的地标挖掘方法
 % 利用了前人未利用的信息来挖掘地标（机构名字）
 % 新的地址提取方法，前人用正则表达式，提高了准确率和地址完整性
% 提高了IP定位的精度

% 缺陷
% 未来方向

%%
%% The acknowledgments section is defined using the "acks" environment
%% (and NOT an unnumbered section). This ensures the proper
%% identification of the section in the article metadata, and the
%% consistent spelling of the heading.
\begin{acks}
This work was supported in part by ...(waiting for updating)

%the National Key R\&D Program of China(Grant No.2018YFB0803402, No.2017YFB0802804), the Key Program of National Natural Science Foundation of China(Grant No.U1766215), and the National Natural Science Foundation of China(Grant No.61702503, No.61702504).
\end{acks}

%%
%% The next two lines define the bibliography style to be used, and
%% the bibliography file.

\bibliographystyle{ACM-Reference-Format}
\bibliography{main}

%%
%% If your work has an appendix, this is the place to put it.
\clearpage
\appendix
\section{LSTM-Ada training}
\subsection{Data Processing}
% 数据的准备
% 部分数据开源
% 如何自动打标签
% 如何缩短文本长度到603
As aforementioned (Section~\ref{sec: exp_clue_extraction}), the dataset WPLICE, which is produced for training and evaluating our model LSTM-Ada, consists of 269,566 web pages. For each sample, the main attributes are organization name, address, and page text. To prepare data for the model, we label the page texts by programming and manual operations according to the address and the organization name. We split the address into 4 items: detailed address, city, state, and ZIP code. Couple with the organization name, we find these 5 entities in the page text and label them to corresponding tags. 

Since LSTM-Ada aims to learn the pattern of the address section and copyright section, we replace the entity tags that outside of these two parts with tag "O". For copyright information, we recognize the section by the character © and only save 100 words around the character ©. Besides, we recognize the address section by calculating the score of the cohesiveness of the 4 items, since these 4 items are ordered in some formats and very close to each other. Most of the page texts consist of tens of thousands of words, which can make the training process too long and waste computing resources, so we shorten the text by cutting off the O-tag words that far away from the entities. More specifically, We leave 100 words before and after the address sequences. After doing this, the maximum of the length of a page text in this dataset is shortened to 603. There can be more than one address sequence and copyright info sequence.

\subsection{Hyper-parameters Setup}\label{sec:hyper_para_model}
% 一些超参设置的补充说明
% 训练策略：先用Adam训练30个epochs，并选择metric最优的模型，再用SGD继续训练10个epochs，选择最佳模型。
We use FastText to pre-train the word embeddings on 22,294,645 web page texts and 11,073,587 addresses. The dimension of the embedding vector is set to 50. For the FastText model, we set to 2 the minimum length of character n-grams and 10 the iterations. As for the model, we set to 50 the dimension of the embedding layer and initialize the layer by the pre-trained word embeddings. Besides, we set to 256 the number of units of the Bi-LSTM encoding layer, 512 the number of units of the Bi-LSTM decoding layer. As for the loss function, we set the coefficient $\alpha$ to 64. We use 4 Titan Xp GPUs to train the model. We set the batch size to 64 and shuffle the data in a buffer of size 120000. During the training process, we first use Adam with default parameters to train the model for 30 epochs and choose the best model by monitoring F1 score of the validation set. Then, we use SGD with a learning rate of 0.001 to continue to train the model for 10 epochs and choose the best one by monitoring the F1 score. 

\section{Determine possible coordinates by the variant of CBG}\label{sec:cbg_var}
% 如果没有任何contact information，用CBG变体确定大概区域，再在此区域搜索organization name
If a web page does not contain any contact information but only an organization name, we decide possible coordinates by a variant of CBG. Mainly, CBG utilizes a set of probes to measure a set of delays to a target IP. Based on the delays and delay-distance relation, CBG draws a set of constraint circles, of which the overlap region is where the target IP resides. Different from the classical CBG, we do not determine the location of the target IP after the intersection region is calculated, instead, we filter the candidate coordinates iteratively. To be more specific, we search candidate coordinates in the smallest circle and then iteratively remove those coordinates that reside outside of the other circles.

\begin{figure}[htb]
    \centering
    \includegraphics[scale=0.37]{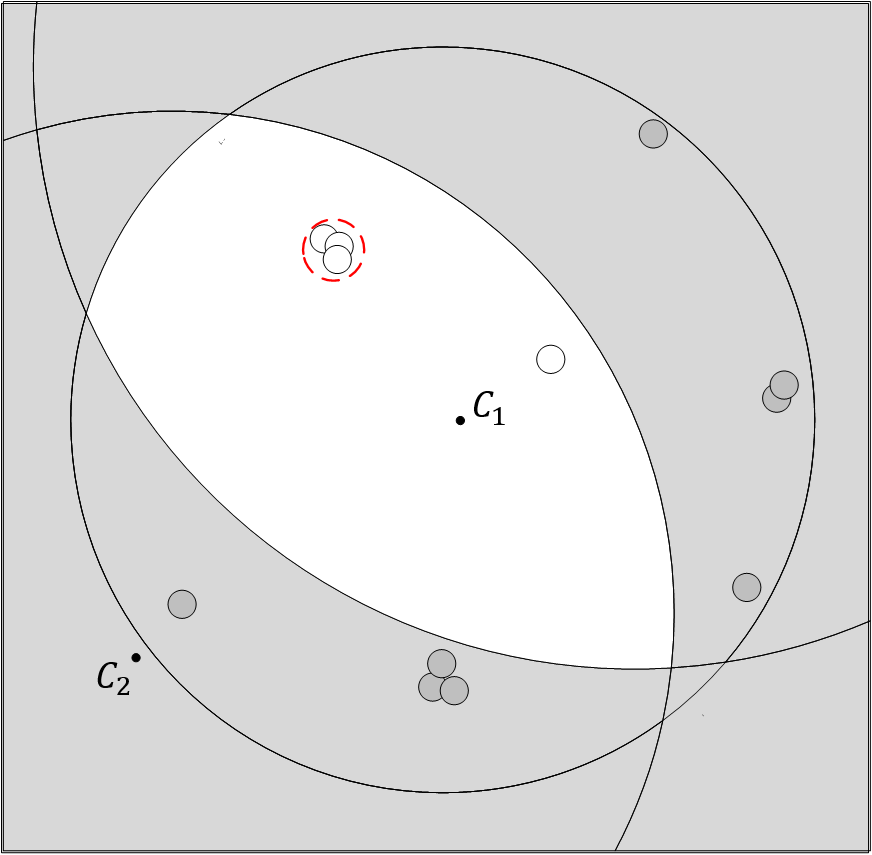}
    \caption{To search candidate locations based on a variant of CBG}
    \label{fig:cbg_candidates}
\vspace{-0.1in}
\end{figure}

% 搜索candidate locations的过程
Figure~\ref{fig:cbg_candidates} illustrates the process. Formally, for each probe $P_i$, we draw a constraint circle centered at the probe, with the estimated distance between the probe and the target as the radius. Let $\{C_1, C_2, ..., C_i, ..., C_n\}$ be the set of circles ordered in ascending order of radius $r_i$. We first query the organization name in $C_1$ by a map service (e.g. Google Map) and get a set of candidate coordinates (small circles in Figure~\ref{fig:cbg_candidates}). Some of them also fall into $C_2$, but the others do not. In the next loop, we remove those coordinates located outside of $C_2$. Similar logic can be applied to the rest loops. In the end, Several coordinates are left in the overlap region (the white region in Figure~\ref{fig:cbg_candidates}), called candidate coordinates. As shown by the red dashed circle, For adjacent coordinates with less than 1 km distance to each other, we merge them into one coordinate. If there is only one candidate coordinate left in the overlap region, we will map it to the target IP directly. If more than one, we use the coordinate selection algorithm (Section~\ref{sec:selection_algorithm}) to select the most possible location.

% 约束圈的半径怎么算
To estimate the radius of a constraint circle, we first use ping to measure the target IP and then convert the delay between each probe and the target device into a geographical distance by Formula~\ref{equ:delay_2_distance}. Percacci et al.\cite{percacci2003scale} has shown that packets travel in fiber optic cables at 2/3 the speed of light in a vacuum (denoted by c) . However, other literature\cite{gueye2006constraint}\cite{tbg} has demonstrated that 2/3 is a loose upper bound of converting factor because it does not take into account the transmission delay, the queuing delay, and the processing delay. Katz-Bassett et al.\cite{tbg} claims 4/9 is a safe threshold for constrains, so we adopt 4/9 as the converting factor $f$. Proved by Wang et al.\cite{slg}, by using this converting factor, the overlap region can always cover the targeted IP.

\begin{equation}
    \label{equ:delay_2_distance}
    r_i = \frac{p_i}{2}\cdot f \cdot c
\end{equation}

\section{Organization dictionary}\label{sec:org_dict}
% 为什么要使用字典
    % LSTM-Ada works well for copyright information due to the regular patterns.
In most cases, LSTM-Ada works well for extracting organization name from copyright information due to the obvious patterns. Nevertheless, for some copyright information with an unconventional format, the model often fails to extract the correct or entire organization name. Besides, for organization names in the page title or the anchor text of logo images, the model also does not work due to the lack of context or regularity. Hence, we build an organization name dictionary to help extract the organization names.

\begin{figure}[htb]
    \centering
    \includegraphics[scale=0.45]{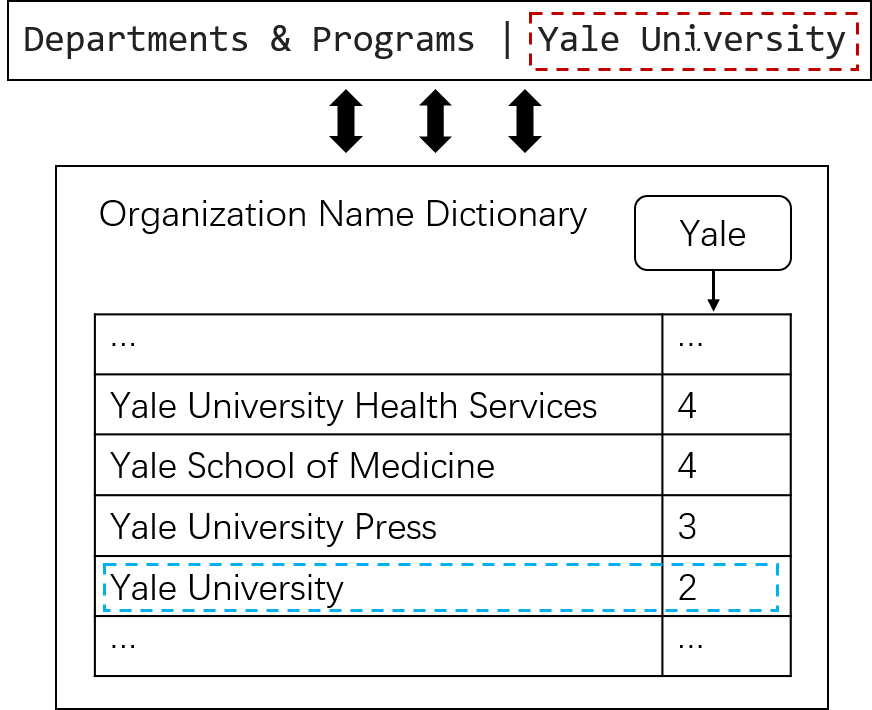}
    \caption{An example of extracting organization name by dictionary}
    \label{fig:example_4_org_dict}
\vspace{-0.1in}
\end{figure}

% 如何获得数据源
Through crawling organization information from yellow pages and collecting POIs through Google Map API, we got the relevant information of 11,310,932 organizations, which contains their names. After removing duplicates, we got 7,757,970 organization names totally. These names are indexed with the first word as the key and the names under the same key are sorted in descending order of the length. With the dictionary, we extract organization from page text following three steps: 1) traverse every word $w_i$ in the text, and take the word $w_i$ as the key to look up the organization names which start with it. 2) traverse every organization name $o_i$ in the returned list, from the start word $w_i$, truncate the word sequence $s$ in the text with the same length as the current organization name $o_i$. 3) if the word sequence $s$ equals the organization name $o_i$, we return it as a result. Figure~\ref{fig:example_4_org_dict} shows an example of the extracting process.
% 词典搜索的算法过程

\section{Open data}
Datasets used in experiments, including WPLICE, PlanetLab-104, WebPages-8M, and RIPE-Atlas-1156 are open to the public for research. Visit https://github.com/131250208/dataset4XLBoost-Geo to download.
\end{document}